\documentclass[12pt]{article}
\usepackage[utf8]{inputenc}

\usepackage{graphicx}
\usepackage{amsmath}
\usepackage{amssymb}
\usepackage{amsfonts}

\usepackage{setspace}
\usepackage[top=0.8in, bottom=0.8in, left=1in, right=1in]{geometry}

\usepackage[]{algorithm}
\usepackage{algpseudocode}

\usepackage[section]{placeins}

\usepackage[
    backend=biber,
    style=chem-rsc,
  ]{biblatex}
\usepackage{mathtools}
\newcommand{\zbf}{\mathbf{z}}
\newcommand{\zt}{\mathbf{z}_t}
\newcommand{\yt}{\mathbf{y}_t}
\DeclarePairedDelimiter{\norm}{\lVert}{\rVert}
\DeclareMathOperator*{\argmin}{arg\,min}

\graphicspath{{./figures/}}

\begin{document}

\title{
Extraction of instantaneous frequencies and amplitudes in nonstationary time-series data}

\author{Daniel E. Shea$^{*}$, Rajiv Giridharagopal$^{**}$, David S. Ginger$^{**}$, \\ Steven L. Brunton$^\dag$,  J. Nathan Kutz$^\ddag$\\
{\small
$^*$ Department of Material Science, University of Washington, Seattle 98195}\\
{\small $^{**}$ Department of Chemistry, University of Washington, Seattle 98195}\\
{\small $^\dag$ Department of Mechanical Engineering, University of Washington, Seattle 98195}\\
{\small $^\ddag$ Department of Applied Mathematics, University of Washington, Seattle 98195}}

\maketitle

\begin{abstract}
\noindent
Time-series analysis is critical for a diversity of applications in science and engineering.   By leveraging the strengths of modern gradient descent algorithms, the Fourier transform, multi-resolution analysis, and Bayesian spectral analysis, we propose a data-driven approach to time-frequency analysis that circumvents many of the shortcomings of classic approaches, including the extraction of nonstationary signals with discontinuities in their behavior.  The method introduced is equivalent to a {\em nonstationary Fourier mode decomposition} (NFMD) for nonstationary and nonlinear temporal signals, allowing for the accurate identification of instantaneous frequencies and their amplitudes.  The method is demonstrated on a diversity of time-series data, including on data from cantilever-based electrostatic force microscopy to quantify the time-dependent evolution of charging dynamics at the nanoscale.
\end{abstract}

\section{Introduction}

Time series data analysis is ubiquitous and foundational in scientific analysis and engineering model design~\cite{hamilton_time_1994}. Indeed, it has revolutionized nearly every scientific discipline by enabling the development of test models for observed natural phenomena in diverse applications that include planetary motion, chemical reactions, meteorological patterns, and transport phenomena. In a typical scientific workflow, observations are made on a system and fit to a time series model, which can include classical methods from statistics, such as ARIMA and its variants~\cite{hamilton_time_1994}, or more recent neural network based approaches~\cite{bengio2017deep,Brunton2019book}, such as LSTM~\cite{hochreiter1997long} (long-term, short-term memory), GRU (gated recurrent units)~\cite{cheRecurrentNeuralNetworks2018}, and echo-state networks~\cite{pathak2018model}.  These diverse mathematical strategies regress to models fit to historical training data, often making assumptions that the data is generated from a stationary process with Gaussian distributed statistics. However, this workflow is often complicated by observations with non-Guassian noise, the existence of nonstationary processes, and/or nonlinear system dynamics.  These challenges make forecasting exceptionally difficult, requiring the re-training of models as new data becomes available.  By integrating elements of modern gradient descent algorithms, the Fourier transform, multi-resolution analysis, and Bayesian spectral analysis~\cite{bretthorst_bayesian_1988}, we can learn an interpretable Fourier mode-based model for analyzing nonstationary signals with periodic components, thus circumventing the challenges normally associated with nonstationary processes and allowing for accurate identification of instantaneous frequencies and their amplitudes.

Joseph Fourier revolutionized time series analysis with the introduction of his eponymous transform in 1822, which he developed while studying heat conduction~\cite{fourier_theorie_1822}. The transform empowered an understanding of the frequency-energy spectrum of time-series signals and spurred the development of Fourier transform-powered time-frequency analyses~\cite{allen_short_1977}. These spectrum-based analysis tools have been extensively applied in systems exhibiting periodic and quasi-periodic behaviors, including oscillators and waves.  So extensive are the applications that the field of {\em harmonic analysis} has emerged as a consequence. Harmonic analysis has been applied to a diverse range of problems spanning many size and time scales, including mechanical vibrations and machine health monitoring~\cite{tu_iterative_2020}, speech and music recognition~\cite{2012HadjidimitriouEEGMusic,2006kepesiTFAspeechmusic}, oceanographic tide modeling~\cite{guoRivertideDynamicsExploration2015}, telecommunications and power systems~\cite{linDevelopmentofanImprovedTime2014}, and quantum mechanics~\cite{griffiths_schroeter_2018}.

Despite its widespread use and generality, the Fourier transform has a number of restrictions that limit its utility in analyzing nonlinear and nonstationary processes~\cite{huang_empirical_1998}. 
This was recognized by Denis Gabor in considering radar technologies of the mid-20th century~\cite{gabor1946theory}.  Indeed, Gabor suggested circumventing these issues in part by using short-time, or windowed, Fourier transforms.  This led to improvements in time-frequency analysis and eventually to the development of wavelet theory~\cite{mallat1999wavelet}.  Gabor and wavelet transforms provide rich visualizations of time-frequency representations with {\em spectrograms} and {\em scalograms} respectively~\cite{kutz2013data}.  More recently, the Fourier transform has inspired a new class of time-frequency methods, detailed below, that complement traditional Fourier transform-based approaches\cite{dragomiretskiy_variational_2014,chen_nonlinear_2017,singh_fourier_2017,kowalski_convex_2018,hirsh2020data,hou2017sparse}. Among them, the Hilbert-Huang transform (HHT)~\cite{huang_empirical_1998} has become a common tool for understanding nonstationary processes. The HHT combines empirical mode decomposition (EMD), which separates a multi-component signal into simpler periodic modes called intrinsic mode functions (IMFs), with Hilbert spectral analysis. Hilbert spectral analysis leverages the Hilbert transform to compute the analytic signal of each periodic mode identified by EMD. In turn, the analytic signal can be leveraged to compute the instantaneous phase, instantaneous frequency, and instantaneous amplitude of the input signal. A significant theoretical framework has been developed around applying the Hilbert transform to vibrational problems~\cite{feldman_hilbert_2011,feldman_non-linear_1994,feldman_non-linear_1994-1}. Although powerful in practice, the HHT is known to have difficulty with signals that exhibit nonlinear phase evolution~\cite{daubechies_synchrosqueezed_2009}. Furthermore, the algorithm behind the EMD is empirical, thus lacking a broader theoretical underpinning. 

Our method is complementary with a newer generation of time-frequency analysis algorithms that address specific shortcomings of classic approaches. The Tycoon method was introduced to handle signals with extremely fast changing frequencies~\cite{kowalski_convex_2018}, but focuses on signals that are (relatively) stationary. The variational mode decomposition (VMD) aims to simultaneously identify periodic modes with time-dependent non-linear phase functions by maximizing the smoothness of the amplitude of the periodic modes~\cite{dragomiretskiy_variational_2014}. VMD has been successfully applied to a broad array of problems, though the importance placed on smooth amplitude functions limits its applications for situations with discontinuities in modes' phase or amplitude functions. A wide array of approaches have been built on the VMD and extend it to different types of signals and multichannel measurements~\cite{rehman_multivariate_2019,chen_nonlinear_2017, tu_iterative_2020}. A different Fourier mode-based algorithm has also been developed which uses an approach similar to basis pursuit~\cite{singh_fourier_2017}. Other approaches focus on decomposition of specific models, such as harmonic oscillators, and the parameters that describe them. These approaches include both Hilbert vibrational decomposition and Kalman filter-based approaches \cite{feldman_non-linear_1994,feldman_non-linear_1994-1,feldman_hilbert_2011,yazdanian_estimation_2015} along with sparsity-promoting decompositions~\cite{hou2017sparse,hirsh2020data}. Wigner distribution-based approaches show great promise for decomposing multicomponent signals with modes that have crossover between frequencies~\cite{stankovic_decomposition_2020}. 

In this work, we develop a method for extracting the instantaneous frequencies and amplitudes from time-series data.  The methods is equivalent to a {\em nonstationary Fourier mode decomposition} (NFMD) for nonstationary and nonlinear temporal signals. Importantly, it produces interpretable signal decompositions that can handle signals with multiple periodic components, non-linear phase functions, and sharp discontinuities in the phase function or periodic mode amplitudes.  Adopting the work of Lange et al~\cite{lange_fourier_2021}, which employed a similar architecture for future state prediction rather than interpretable time-frequency analysis, the proposed method leverages modern gradient descent optimization to fit temporally-local linear Fourier modes. The approach resembles the {\em short time Fourier transform} (STFT) wherein smaller temporal segments of the signal are analyzed independently, and the resulting analyses are combined to provide a full time-frequency representation of the signal. The NFMD fits Fourier modes to each signal segment, and computes the mode frequency and amplitude for each signal segment through a gradient descent optimization with a nonlinear Fourier basis objective function. The method results in a superior time-frequency analysis to the HHT for nonstationary signals, and improves both temporal and spatial resolutions compared to the STFT. The NFMD can be applied to systems with fast-changing frequencies and abrupt changes in the signal mean, such as machine health monitoring, seismology, vibration-based imaging modalities, neurochemical and biochemicals signals, and photonic sensing.

\section{Methods}

The NFMD analysis proposed combines elements of modern gradient descent algorithms, the Fourier transform, Bayesian spectral analysis, and an algorithm similar to STFT to learn an interpretable Fourier mode-based model for analyzing nonstationary signals with periodic components. In general, we aim to fit a model $\yt$ to a measured signal $\zt$. We begin by framing the Fourier mode decomposition approach for an entire time series signal. The subsequent section shows how the method is combined with a segment-by-segment analysis, reminiscent of STFT, to propose a full time-frequency analysis. Finally, the instantaneous signal parameters and nonstationary part of the signal are addressed. We present a simple algorithm for computing the nonstationary signal component from the learned Fourier mode representations.
 
\subsection{Fourier Mode Decomposition}\label{sec:method-fmd}

Consider the time series data $\zt[\zbf(t_1),\zbf(t_2),\cdots,\zbf(t_n)]$ sampled at $n$ discrete evenly-spaced times in $t \in \mathbb{R}^+$. The signal is assumed to be periodic or quasi-periodic. The most common frequency-domain signal analysis method is the Fourier transform, which allows the input signal $\zt$ to be represented as a Fourier series. The series representation is a sum of sines and cosines that provides insight into the frequency-energy spectrum of the signal. 

The Fourier transform, typically implemented as the Fast Fourier Transform (FFT), assumes a periodic input signal that satisfies the relationship $\zbf(t_i) = \zbf(t_i+P)$, where $P$ is the period of the periodic data. This periodicity assumption imparts a limited frequency resolution, $F_s/P$, that scales linearly with the sampling frequency of the signal, $F_s$, and inversely with the period $P$. Note the period $P$ is typically set to the total time $(t_n-t_1)$ of the discrete input signal.

\begin{algorithm}[t]
\caption{\label{fmd-algo}Fourier mode signal decomposition (FMD)}
\hspace*{\algorithmicindent} \textbf{Input:}  Signal $\zt$, Initial frequency guess $\boldsymbol{\omega}$, Error tolerance $tol$ \\
\hspace*{\algorithmicindent} \textbf{Output:} Learned frequency $\boldsymbol{\omega}$, Learned amplitude $\mathbf{A}$, Residual error $E(\mathbf{A},\boldsymbol{\omega})$ \\
	\begin{algorithmic}[1]
		\Procedure{}{}
			\State $\mathbf{A} \gets \zt (\Omega(\boldsymbol{\omega} t))^{-1}$ \Comment{Amplitude from initial frequency guess}
			\While{$E(\mathbf{A}, \boldsymbol{\omega}) > tol$} \Comment{Use gradient descent to optimize frequency vector $\boldsymbol{\omega}$}
				\State $\boldsymbol{\omega} \gets \argmin_{\boldsymbol{\omega}^*} E(A, \boldsymbol{\omega}^*)$
				\State $\mathbf{A} \gets \zt (\Omega(\boldsymbol{\omega} t))^{-1}$
				\State Update $E(\mathbf{A},\boldsymbol{\omega})$
			\EndWhile \label{FMD Loop}
			\State \textbf{return} $\boldsymbol{\omega}$, $A$, $E(A, \boldsymbol{\omega})$
		\EndProcedure
	\end{algorithmic}
\end{algorithm}

To overcome these limitations of the FFT, we adapt the Fourier series representation and exploit some of its mathematical properties to enable the NFMD time series analysis. First, the Fourier series model is framed in a more general context which uses a finite number of modes and flexible mode frequencies which need to be determined. The model
\begin{equation*}
    \yt(t) = \yt = \sum_{k=1}^{K} F_k(t)= \sum_{k=1}^{K} a_k \cos(\omega_k t) + b_k \sin(\omega_k t)
\end{equation*}
is used, where $F_k(t)$ is the general Fourier mode function, the coefficients $a_k$ and $b_k$ are the weights for the cosine and sine components of each mode, respectively, $\omega_k$ is the frequency of mode $k$, and $K$ modes are considered. In matrix form, the model is
\begin{equation}
    \mathbf{y}_t(t) = \mathbf{A} \Omega(\boldsymbol{\omega} t) = 
    \begin{bmatrix}
        a_1 & \dots & a_K & b_1 & \dots & b_K
    \end{bmatrix}
    \begin{bmatrix}
        \cos(\omega_1 t) \\ \vdots \\ \cos(\omega_K t) \\ \sin(\omega_1 t) \\ \vdots \\ \sin(\omega_K t)
    \end{bmatrix},
\end{equation}
where $\mathbf{A} \in \mathbb{R}^{2\times K}$ is a vector of the coefficients, and $\Omega(\boldsymbol{\omega} t) \in \mathbb{R}^{2\times K}$ is a vector of cosines and sines with frequency $\boldsymbol{\omega}\in \mathbb{R}^K$. The vector $\mathbf{A}$ is determined as the optimal coefficient vector given a frequency vector $\boldsymbol{\omega}^*$, by fitting to the time-series data $\zt$, using the computation
\begin{equation*}
    \mathbf{A} = \zt (\Omega(\boldsymbol{\omega}^* t))^{-1}.
\end{equation*}
This is a common approach for discovering component amplitudes in Bayesian spectral analysis~\cite{bretthorst_bayesian_1988}. The vector $\boldsymbol{\omega} = [\omega_1, \dots, \omega_K]$ is determined by the optimization 
\begin{equation}
    \textnormal{minimize} \quad E(A, \boldsymbol{\omega}) = \sum_{t \in T}(\zt - A \Omega(\boldsymbol{\omega}t))^2.
\end{equation}
This method has been previously demonstrated, including the theory behind the optimization \cite{lange_fourier_2021}. This approach does \emph{not} appear to be a convex optimization objective, given the nonlinear objective with cosine and sine functions in $\Omega(\cdot)$, and therefore should not yield globally optimal solutions. Although this objective does lack global convexity, this pitfall is avoided by using initial guesses near the optimal solutions by leveraging the FFT for an initial guess. The initial guess frames this problem on an error surface that is locally convex, and therefore allows for highly accurate frequency estimation by gradient descent. Ultimately, this allows our input signal to be fit to a model that is a superposition of Fourier modes with superior frequency resolution to traditional Fourier transforms. The {\em Fourier mode decomposition} (FMD) algorithm for decomposing a full time series signal $\zt$ is presented in Algorithm \ref{fmd-algo}. This algorithm enables the time-frequency analysis framework described in the next section.

\subsection{Nonstationary Fourier Mode Decomposition}

\begin{figure}[t]
    \centering
    \includegraphics[width=15cm]{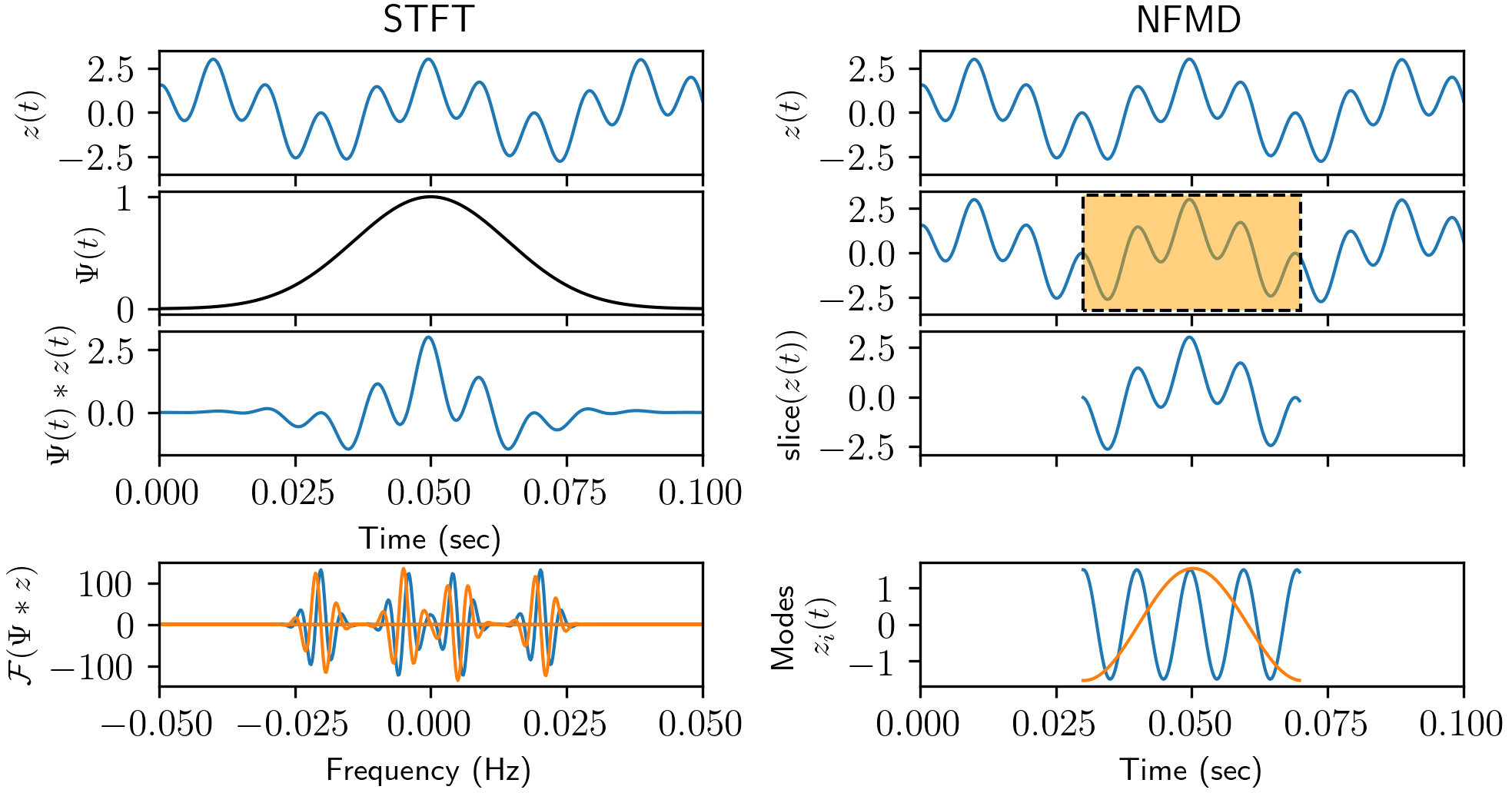}
    \caption{\textbf{Comparison between STFT and NFMD methods.} The STFT computes the Fourier transform of a convolution of the signal and a windowing function, $\Psi(t)$. The example shown uses a Gaussian windowing function, which is commonly referred to as the Gabor transform. The bottom left panel shows the real (blue) and imaginary (orange) components of the Fourier transform. The NFMD analyzes a specific segment of the signal, and fits a finite number of Fourier modes to the signal segment. The modes fit to the signal segment are shown in the bottom right panel, colored blue and orange.}
    \label{fig:nfmd-vs-stft}
\end{figure}

The NFMD algorithm builds upon Fourier decompositions to create a descriptive time-frequency analysis (TFA) framework that exhibits improved frequency resolution compared to traditional TFA techniques. The NFMD is similar in principle to one of the most common traditional techniques for TFA, the STFT. The STFT determines the frequency-domain energy spectrum of temporally local intervals of a signal. Although effective, the reliance on the traditional Fourier transform technique limits the frequency resolution of the STFT and has limited interpretability for understanding the signal components. Figure \ref{fig:nfmd-vs-stft} presents a graphical comparison between the time-frequency analysis approach of the STFT and the NFMD algorithm. In the STFT, the original signal is multiplied by a set of windowing functions, such as a Gaussian, that is progressively applied to subsets of the signal. The signal subsets are then analyzed by the FFT. In the NFMD, signal segments are sliced and analyzed individually; the resultant Fourier modes are then fit by the FMD algorithm described in Section \ref{sec:method-fmd}, thereby enabling a time-frequency analysis.

Specifically, NFMD subsamples segments of the time series of length $\xi$. These segments take the form $\boldsymbol{\chi}_{i}=\{{\bf z}(t_{i}),{\bf z}(t_{i+1})...,{\bf z}(t_{i+\xi/2})\}$. There are $n - \xi$ segments considered for an input signal of length $n$. The set of all segments in the signal is the set ${\bf X} = \{\boldsymbol{\chi}_1, \boldsymbol{\chi}_2, ..., \boldsymbol{\chi}_i, ..., \boldsymbol{\chi}_{n-\xi} \}$. The NFMD algorithm is presented in Algorithm \ref{nfmd-algo}.

\begin{algorithm}[t]
\caption{\label{nfmd-algo}Nonstationary Fourier Mode Decomposition}
\hspace*{\algorithmicindent} \textbf{Input:}  Signal segments $X$, Window size $\xi$, tolerance $tol$ \\
\hspace*{\algorithmicindent} \textbf{Output:} Frequency matrix, $\hat{\omega}$, Coefficients matrix $\hat{\mathbf{A}}$, residual error vector $\hat{E}$
	\begin{algorithmic}[1]
		\Procedure{}{}
		    \State $\boldsymbol{\omega}_{-1} \gets$ maxima(FFT($\boldsymbol{\chi}_1$)) \Comment{Initial guess for frequencies using the FFT}
			\For{$\boldsymbol{\chi}_i \; \in \; X$}
				\State $\mathbf{A}_i$, $\boldsymbol{\omega}_i$, $E_i$ = FMD($\boldsymbol{\chi}_i$, $\boldsymbol{\omega}_{-1}$, $tol$)	\Comment{Learn Fourier modes for segment $\boldsymbol{\chi}_i$}
				\State $\boldsymbol{\omega}_{-1} \gets \boldsymbol{\omega}_{i}$
			\EndFor\label{NFMD Loop}
			\State $\hat{\mathbf{A}} = [\mathbf{A}_1^T, \cdots, \mathbf{A}_i^T, \cdots, \mathbf{A}_{n-\xi}^T]^T$ \Comment{Coefficients matrix}
	    	\State $\hat{\boldsymbol{\omega}} = [\boldsymbol{\omega}_1^T, \cdots, \boldsymbol{\omega}_i^T, \cdots, \boldsymbol{\omega}_{n-\xi}^T]^T$ \Comment{Frequencies matrix}
	    	\State $\hat{\mathbf{E}} = [E_1, \cdots, E_i, \cdots, E_{n-\xi}]$ \Comment{Residual errors vector}
			\State \textbf{return} $\hat{\mathbf{A}}$, $\hat{\boldsymbol{\omega}}$, $\hat{\mathbf{E}}$
		\EndProcedure
	\end{algorithmic}
\end{algorithm}

The NFMD algorithm takes in a set, ${\bf X}$, of window segments, $\boldsymbol{\chi}_i$, and for each $\boldsymbol{\chi}_i \in X$ learns a coefficient vector ${\bf A}_i$ and frequency vector $\boldsymbol{\omega}_i$ using FMD (Algorithm \ref{fmd-algo}). Importantly, the learned frequency vector from each segment is used as the initial guess to the subsequent segment. This significantly increased the speed of the algorithm, by allowing each segment to start with a nominally close estimate of the frequency vector $\boldsymbol{\omega_i}$. The learned vectors from all of the segments are stored in matrices 
\begin{equation}
    \hat{{\bf A}} = \begin{bmatrix}
        - \; {\mathbf{A}_{1}} \; - \\
        \vdots \\
        - \; {\mathbf{A}_i} \; - \\
        \vdots \\
        - \; {\mathbf{A}_{n-\xi}} \; -
    \end{bmatrix},
\end{equation}
and 
\begin{equation}
    \hat{\boldsymbol{\omega}} = \begin{bmatrix}
        - \; \boldsymbol{\omega}_1 \; - \\
        \vdots  \\
        - \; \boldsymbol{\omega}_i \; - \\
        \vdots \\
        - \; \boldsymbol{\omega}_{n-\xi} \; - 
    \end{bmatrix}.
\end{equation}
The learned frequency vectors, $\boldsymbol{\omega}_i = [\omega_{1,i}, ..., \omega_{k,i}, ... \omega_{K,i}]$, and coefficient vectors, $\mathbf{A}_i = [A_{1,i}, ..., A_{k,i}, ..., A_{K,i}]$, contain the frequencies and coefficients for the $K$ fourier modes of each signal segment $\boldsymbol{\chi}_i \in X$.

\subsection{Instantaneous Frequency and Amplitude}

The NFMD decomposes nonstationary, nonlinear signals by fitting a set of Fourier modes to a set ${\bf X}$ of signal segments, as previously described. Importantly, for each signal segment $\boldsymbol{\chi}_i$ the algorithm yields a frequencies vector $\boldsymbol{\omega}_i$ and coefficients vector $\mathbf{A}_i$. For each Fourier mode $k$, a vector of instantaneous frequencies can be constructed by collecting the learned Fourier mode frequency $\omega_{k,i}$ for each of the signal segments $i$ for a given mode $k$. The instantaneous frequency vector takes the form
\begin{equation*}
    \boldsymbol{\omega}_k = [\omega_{k,1}, ..., \omega_{k,i}, ..., \omega_{k,n-\xi}].
\end{equation*}
The instantaneous amplitude of a Fourier mode can be computed from its coefficients $a_{k,i}$ and $b_{k,i}$ with the relationship
\begin{equation*}
    \phi_{k,i} = \sqrt{a_{k,i}^2 + b_{k,i}^2}.
\end{equation*}
Similar to the instantaneous frequency vector, an instantaneous amplitude vector can be constructed for each mode $k$ with one element corresponding to each segment $\boldsymbol{\chi}$,
\begin{equation*}
    \boldsymbol{\phi}_{k} = [\phi_{k,1}, ..., \phi_{k,i}, ..., \phi_{k,n-\xi}].
\end{equation*}
These metrics are useful for comparing the time-frequency analysis from NFMD to other existing time-frequency analysis methods that report instantaneous amplitude and instantaneous frequency of an input time series signal.

\subsection{Nonstationary Signals}
NFMD can provide insight into nonstationary signals of the form 
\begin{equation}
    z(t) = \mu(t) + \sum_{l=1}^{L} A_l(t) \cos(\phi_l(t)),
\end{equation}
where $\mu(t)$ is an unknown, non-periodic function describing the nonstationary part of the signal and there are $L$ periodic cosine modes with the amplitude function $A_l(t)$ and instantaneous phase function $\phi_l(t)$. The term $\mu(t)$ is only assumed to be continuous. We will call the function $\mu(t)$ the signal's \emph{instantaneous mean}. The concept of an instantaneous mean will prove useful in the analysis of nonstationary signals where the signal mean drifts or trends away from zero.

The frequency and coefficient vectors allow computation of individual Fourier modes $F_{k,i} = a_{k,i} \cos(\omega_{k,i} t) + b_{k,i} \cos(\omega_{k,i} t)$ for each of the modes $k \in [1,K]$ and signal segments $\boldsymbol{\chi}_i$. In signals that exhibit a moving mean, there is a mode that will account for the nonstationary part of the signal. The instantaneous mean of the signal, $\mu(t)$, can be estimated from the Fourier modes $F_{k,i}$ corresponding to the mode that accounts for the nonstationary part of the signal. Many interpolation strategies can be implemented to compute the instantaneous mean. We implement a simple strategy of concatenating the value of the Fourier mode $F_k$ at the median time from each time segment $\boldsymbol{\chi}_i$. To visualize this implementation, imagine we construct the matrix
\begin{equation*}
    \mathcal{M} = \begin{bmatrix}
        - \; F_{k,1} \; - \\
        \vdots \\
        - \; F_{k,i} \; - \\
        \vdots \\
        - \; F_{k,n-\xi} \; -
    \end{bmatrix} = \begin{bmatrix}
        F_{k,1}(t_1) & \dots & F_{k,1}(t_{1+\xi/2}) & \dots & F_{k,1}(t_{1+\xi}) \\
        \vdots & & \vdots & & \vdots \\
        F_{k,i}(t_i) & \dots & F_{k,i}(t_{i+\xi/2}) & \dots & F_{k,i}(t_{i+\xi}) \\
        \vdots & & \vdots & & \vdots \\
        F_{k,n-\xi}(t_{n-\xi}) & \dots & F_{k,n-\xi}(t_{n-\xi/2}) & \dots & F_{k,n-\xi}(t_{n})
    \end{bmatrix}.
\end{equation*}
The approach is to use the center column of the matrix $\mathcal{M}$ as the instantaneous mean. This mean is defined as
\begin{equation*}
    \boldsymbol{\mu} = [F_{k,1}(t_{1+\xi/2}), ..., F_{k,i}(t_{i+\xi/2}), ..., F_{k,n-\xi}(t_{n-\xi/2})].
\end{equation*}
The mode representing the mean can be challenging to identify, and generally requires inspection of each of the modes identified by NFMD. In this work, the instantaneous mean is always the lowest-frequency mode.
\section{Results}
We benchmark the NFMD against the Hilbert-Huang Transform for time-frequency analysis of a series of nonstationary multi-component signals. The two methods are compared with and without noise, and on signals with abrupt changes in instantaneous frequency and instantaneous amplitude. After comparing NFMD with the HHT, a pair of oscillator examples are provided to demonstrate how the NFMD and the discovered instantaneous mean can provide insight into the forcing function applied to the oscillator. Finally, the method is applied to simulated and experimental data for a real-world oscillator-based microscopy method. The instantaneous mean is proven to be effective in an experimental data set by validating the form of the discovered instantaneous mean against experimental control data with known forcing functions.
\subsection{Synthetic Test Signals}
\begin{figure}[t]
    \centering
    \includegraphics[width=15cm]{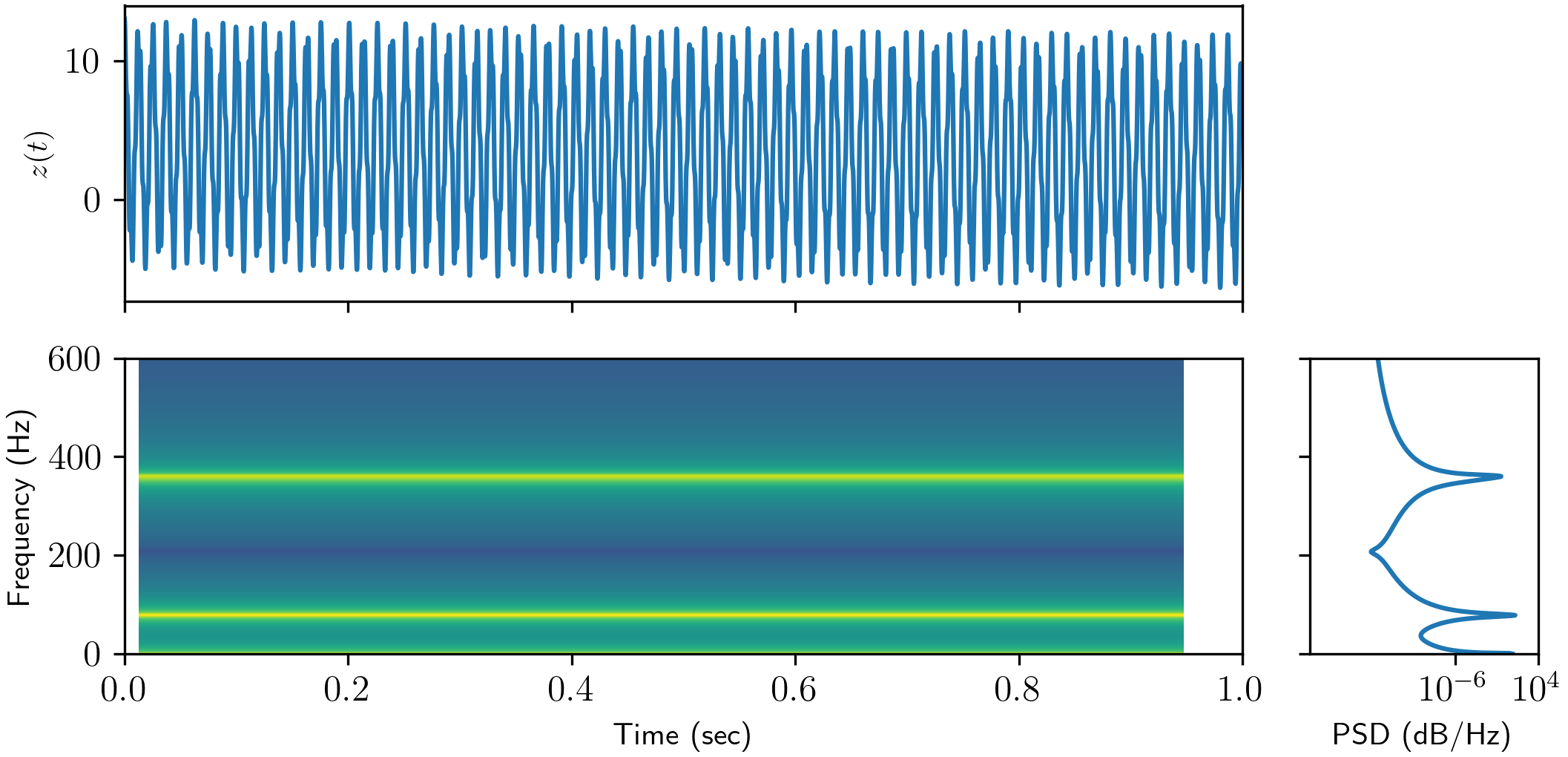}
    \caption{\textbf{Example signal for demonstrating NFMD.} The true signal is presented with the short time Fourier transform and the power spectral density (PSD). The signal has three maxima that appear in the PSD and STFT, indicating that there are likely three modes to be considered.}
    \label{fig:example-stft-psd}
\end{figure}

\begin{figure}[t]
    \centering
    \includegraphics[width=15cm]{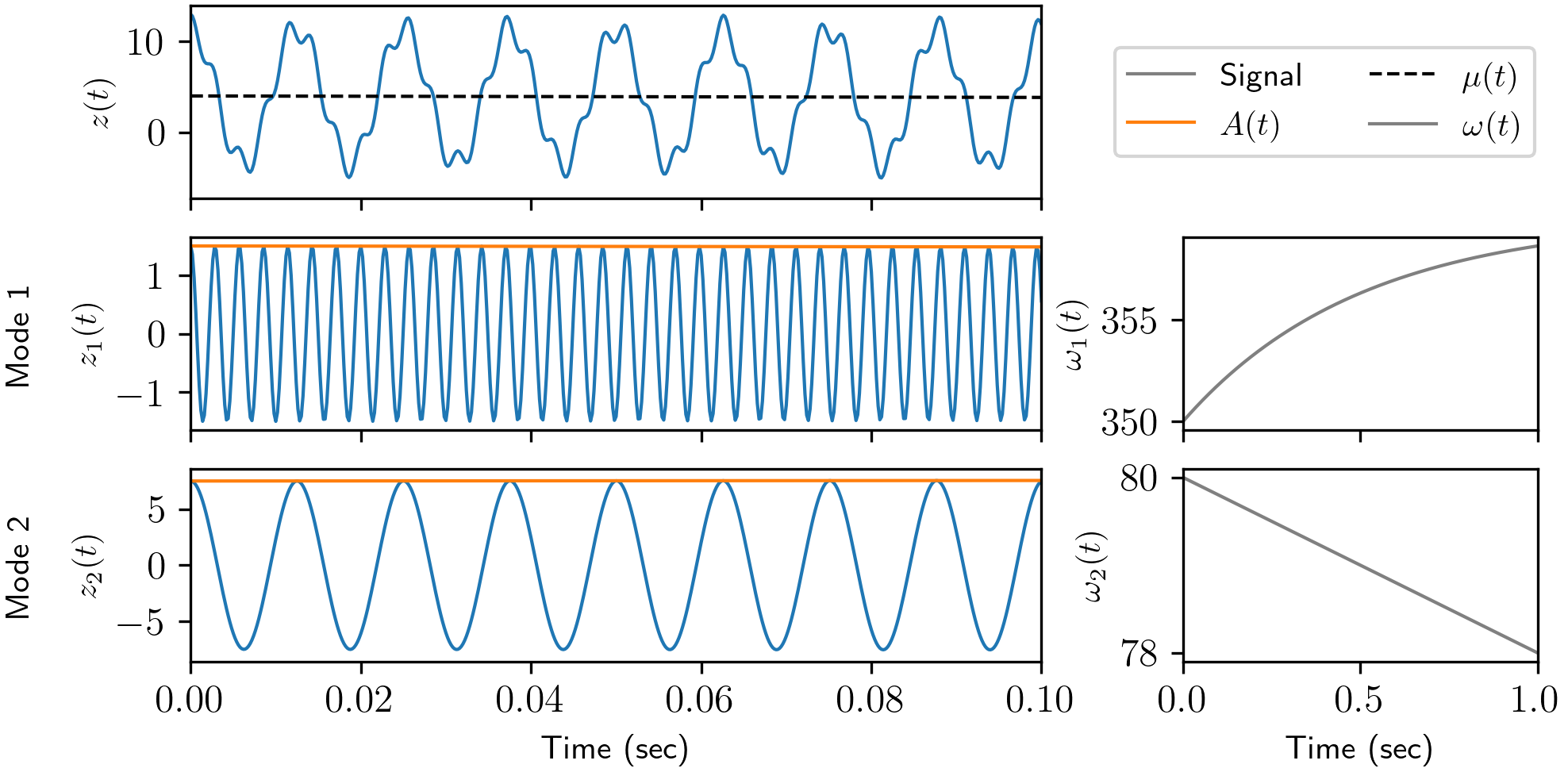}
    \caption{\textbf{Instantaneous mean, amplitude, and frequency} of the example signal. Only part of the signal, from $t=0$ to $t=0.1$ is shown. The instantaneous frequency, $\omega(t)$, and amplitude, $A(t)$, are presented for the two periodic modes. The mean, $\mu(t)$ of the signal is denoted on the signal itself.}
    \label{fig:example-modes}
\end{figure}

Consider a basic multi-component signal with a non-periodic mean:
\begin{align*}
    z(t) &= z_1(t) + z_2(t) + \mu(t) \\
    z_1(t) &= A_1(t) \cos(2 \pi \omega_1(t) t) \\
    z_2(t) &= A_2(t) \cos(2 \pi \omega_2(t) t),
\end{align*}
with amplitude, phase, and instantaneous mean functions
\begin{align*}
    A_1(t) &= 1 + 0.5 \exp (-t/3) \\
    \omega_1(t) &= 360 - 10 \exp(-t/0.5) \\
    A_2(t) &= 8-0.5 \exp(-t) \\
    \omega_2(t) &= 80-2t \\
    \mu(t) &= 1.5 + 2.5 \exp (-x/1.5),
\end{align*}
where $z(t)$ is the signal, $z_1(t)$ and $z_2(t)$ are the periodic components in $z(t)$, and $\mu(t)$ is the slow-moving non-periodic mean. Figure \ref{fig:example-stft-psd} shows a traditional STFT spectrogram and PSD analysis of the signal, and Figure \ref{fig:example-modes} shows the true periodic modes, with instantaneous parameters, and the mean $\mu(t)$. The signal is generated over one second with sampling interval $\Delta t=2 \times 10^{-4}$ s.

\begin{figure}[t]
    \centering
    \includegraphics[width=15cm]{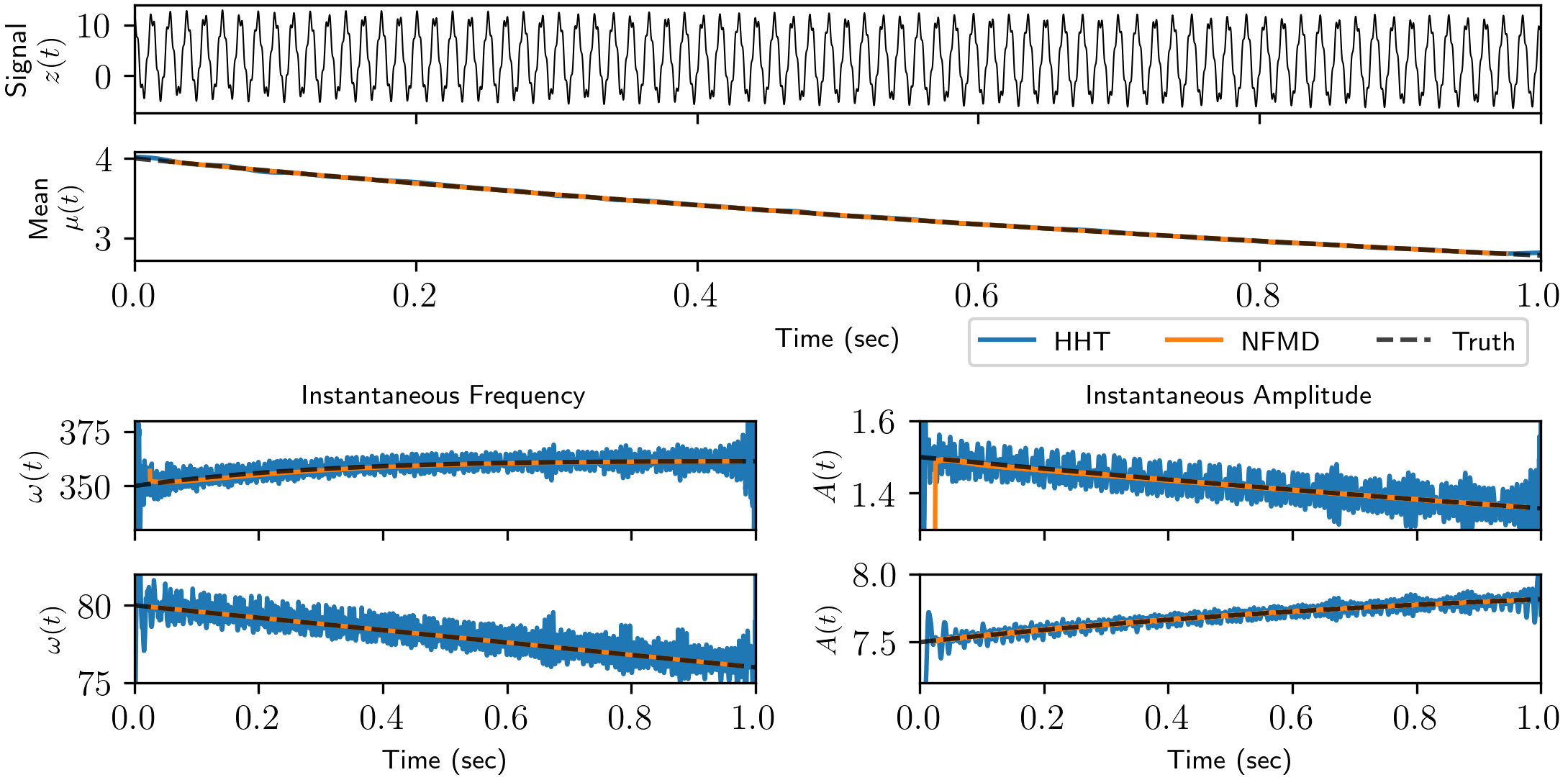}
    \caption{\textbf{Decomposition of example signal by NFMD and HHT without noise.} The estimated instantaneous mean mode is shown for both the HHT and NFMD decomposition methods. The instantaneous frequency and amplitude of the two periodic components is also shown for both decomposition approaches.}
    \label{fig:example-clean-compare}
\end{figure}

\begin{figure}[t]
    \centering
    \includegraphics[width=15cm]{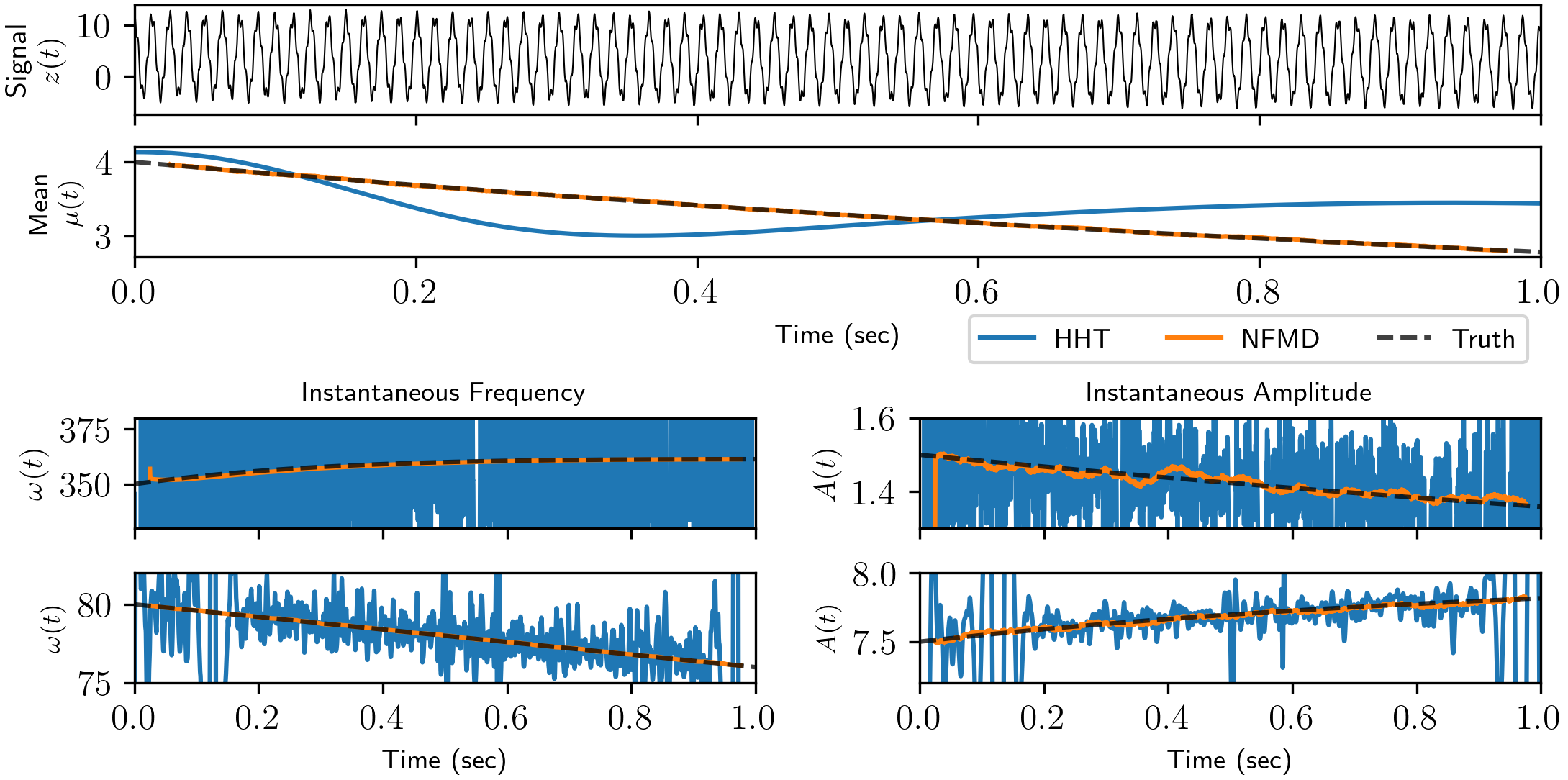}
    \caption{\textbf{Decomposition of example signal with added noise (SNR=35).} The NFMD correctly identifies the instantaneous frequency, amplitude, and mean of the signal, although it does propagate noise in the system. The HHT begins to show erratic behavior in its estimates of all instantaneous parameters.}
    \label{fig:example-noise-compare}
\end{figure}

Signal decomposition is performed with both the NFMD and HHT. The HHT employs the EMD to identify a set of empirical modes. The EMD takes a number of hyperparameters which adjust the outputs, including the number of identified modes. The HHT used in this work used $\theta_1=0.05$, $\theta_2=0.5$, $\alpha=0.05$, and identified four modes in the signal. The two most important hyperparameters for the NFMD are the window size (similar to STFT) and the number of modes to fit to the data. For this data, three modes are fit to a window size of 250 points (or $0.05$ seconds). Figure \ref{fig:example-clean-compare} shows the decomposed signal for both HHT and NFMD, where NFMD does a significantly better job of identifying both the instantaneous frequency and instantaneous amplitude of the signal. Adding noise to the signal significantly affects the HHT's decomposition when compared to the NFMD, as demonstrated in Figure \ref{fig:example-noise-compare}. Noise in the signal is added at a signal-to-noise ratio (SNR) of 35. The SNR in this work is SNR $= 10 \log_{10}(\norm{u(t)}_2^2/\norm{(\tilde{u}(t)-u(t)}_2^2)$.

\begin{figure}[t]
    \centering
    \includegraphics[width=15cm]{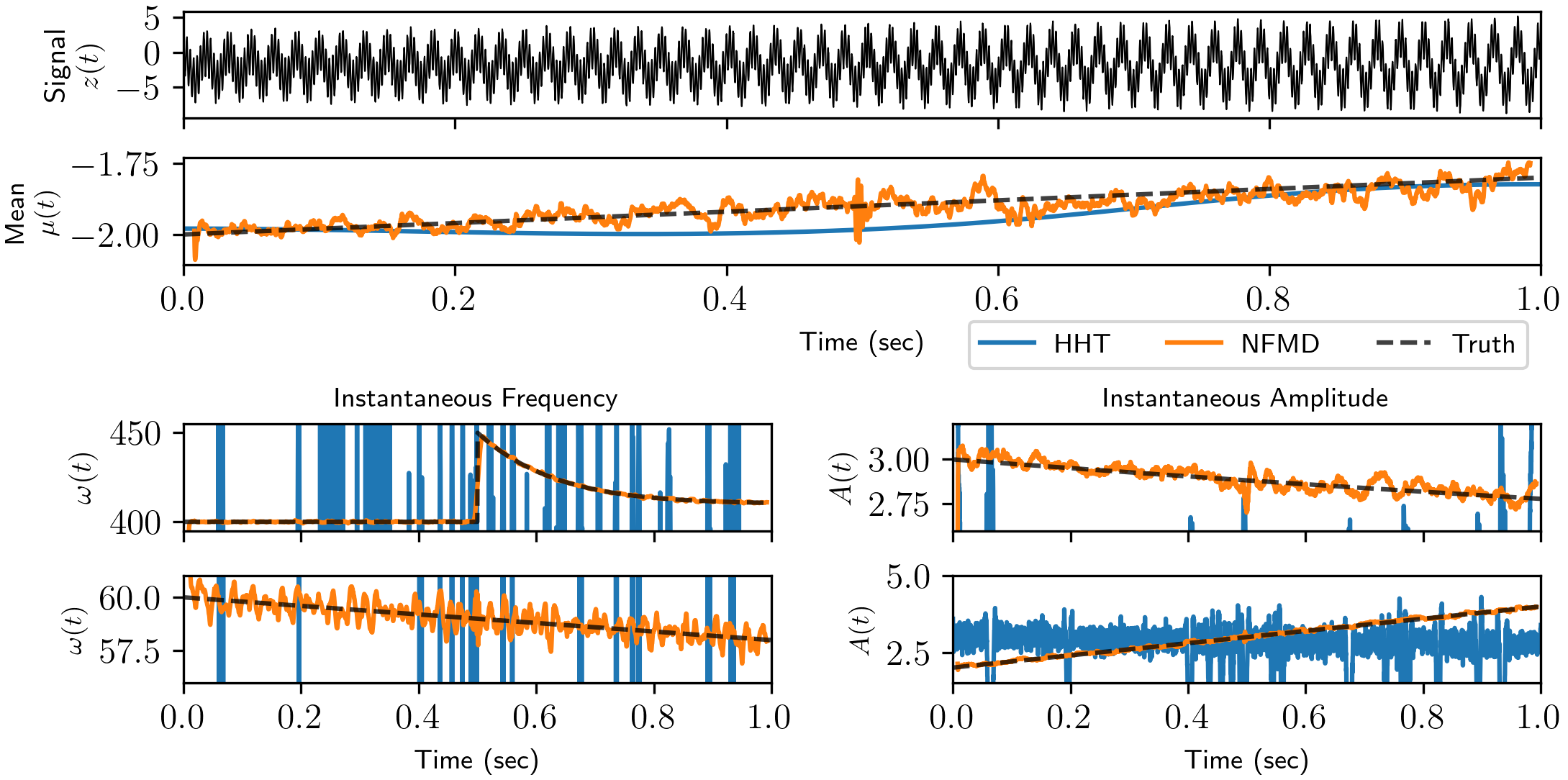}
    \caption{\textbf{Decomposition of signal with discontinuity in instantaneous frequency of a periodic mode.} Data contains additive white noise with SNR=25. The NFMD correctly identifies the three modes, including the instantaneous mean and the frequency and amplitudes of the periodic modes. The HHT fails to identify the correct number of modes and assigns extra modes to the data. The periodic modes uncovered by HHT do not qualitatively match any of the correct signal modes.}
    \label{fig:sharp-omega}
\end{figure}
NFMD enables signal decomposition even for noisy signals where the HHT begins to have difficulty identifying intrinsic mode functions. The NFMD propagates the noise from the signal through the decomposition process, showing some noise-like error in the estimated instantaneous frequency vectors, amplitude vectors, and mean. However, the NFMD provides clearly superior decomposition of the signal.

One particular advantage of NFMD is the ability to correctly identify sharply-changing instantaneous frequency in periodic modes or abrupt changes in the instantaneous mean of an input signal. This is an advantage against other modern optimization-based methods built on VMD, because VMD prioritizes finding solutions with smooth amplitude functions~\cite{dragomiretskiy_variational_2014,tu_iterative_2020}. However, the NFMD has no preference for smooth amplitudes and can accurately decompose these types of signals with abrupt changes in the instantaneous frequency, instantaneous amplitude, or mean. For example, consider a signal with an abrupt change in the instantaneous frequency of one of its modes at $t=0.5$. The model for this example is
\begin{align*}
    z(t) &= z_1(t) + z_2(t) + \mu(t) \\
    z_1(t) &= A_1(t) \cos(2 \pi \omega_1(t) t) \\
    z_2(t) &= A_2(t) \cos(2 \pi \omega_2(t) t),
\end{align*}
with amplitude, phase, and instantaneous mean functions
\begin{align*}
    A_1(t) &= 2 + \exp (-t/4) \\
    \omega_1(t) &= 400 + 10 H(0.5) (1-\exp ((t-0.5)/0.1) \\
    A_2(t) &= 2t + 2 \\
    \omega_2(t) &= 60 - t \\
    \mu(t) &= 1.5 + 2.5 \exp (-x/1.5),
\end{align*}
where $H(0.5)$ is the Heaviside function centered at $t=0.5$s. Noise is added to the signal at an SNR ratio of 25 and decomposed by both HHT and NFMD. The decomposed signal is shown in Figure \ref{fig:sharp-omega}. The NFMD correctly identifies the sharp transition in frequency, and consequently obtains qualitatively good estimates on the amplitude of both periodic modes. Results from the HHT are presented, but it is important to note that in cases like this example, the HHT often yields extra intrinsic mode functions. It is challenging to assign which of the HHT modes are intended to represent which periodic mode, so the presented results were quantitatively closest to the 'true' mode by comparing both instantaneous frequency and instantaneous amplitude vectors. Additionally, all of the remaining modes from HHT are summed together to estimate the instantaneous mean. Although the HHT yields a reasonable looking instantaneous mean, neither of the periodic modes are correctly decomposed in this example.

\begin{figure}[t]
    \centering
    \includegraphics[width=15cm]{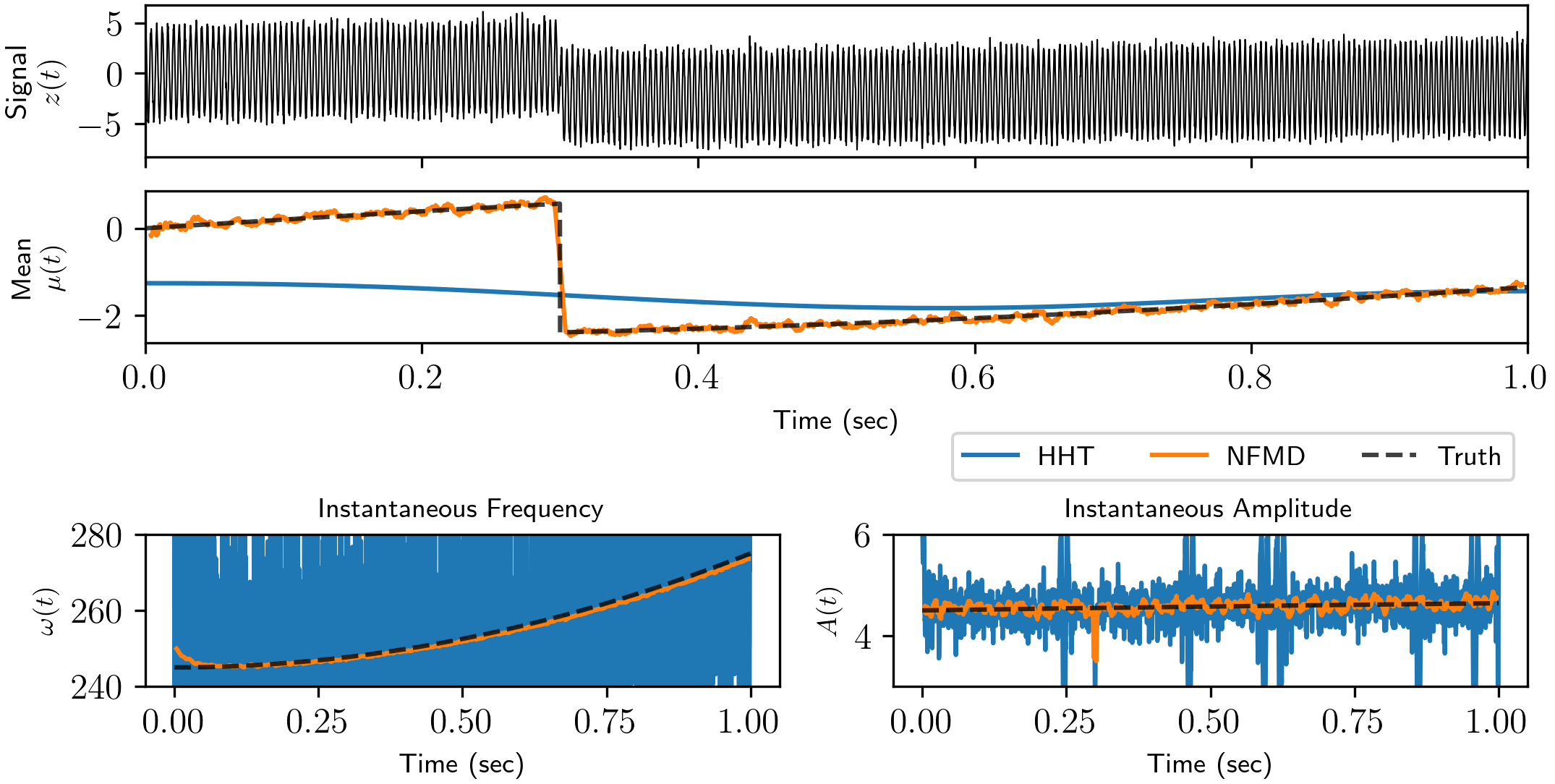}
    \vspace*{-.1in}
    \caption{\textbf{Decomposition of signal with discontinuity in instantaneous mean.} Data contains additive white noise with SNR=20. The NFMD correctly identifies the periodic mode instantaneous frequency and amplitude, and the instantaneous mean of the signal. The HHT fails to identify either the mean or the periodic component.}
    \label{fig:sharp-mean}
\end{figure}

As another example, we consider a situation where the instantaneous mean of the signal changes abruptly in the middle of the signal ($t=0.5$ seconds). The input signal uses the model
\begin{align*}
    z(t) &= z_1(t) + \mu(t) \\
    z_1(t) &= A_1(t) \cos(2 \pi \omega_1(t) t) 
\end{align*}
with amplitude, phase, and instantaneous mean functions
\begin{align*}
    A_1(t) &= 5 - 0.5 \exp (-t/3) \\
    \omega_1(t) &= 245 + 10t^2 \\
    \mu(t) &= \left \{
        \begin{array}{ll}
            \sin{}(2t) & \quad t \leq 0.25 \\
            -2.5 \cos{}(t) & \quad t > 0.25.
        \end{array} \right.
\end{align*}
The signal has added Gaussian noise with SNR$=20$. As shown in Figure \ref{fig:sharp-mean}, the NFMD accurately estimates instantaneous signal mean and identifies the correct periodic mode. The HHT fails to find the instantaneous frequency of the periodic mode, and errantly suggests a low-frequency wave as the instantaneous mean.

\begin{figure}[t]
    \centering
    \includegraphics[width=15cm]{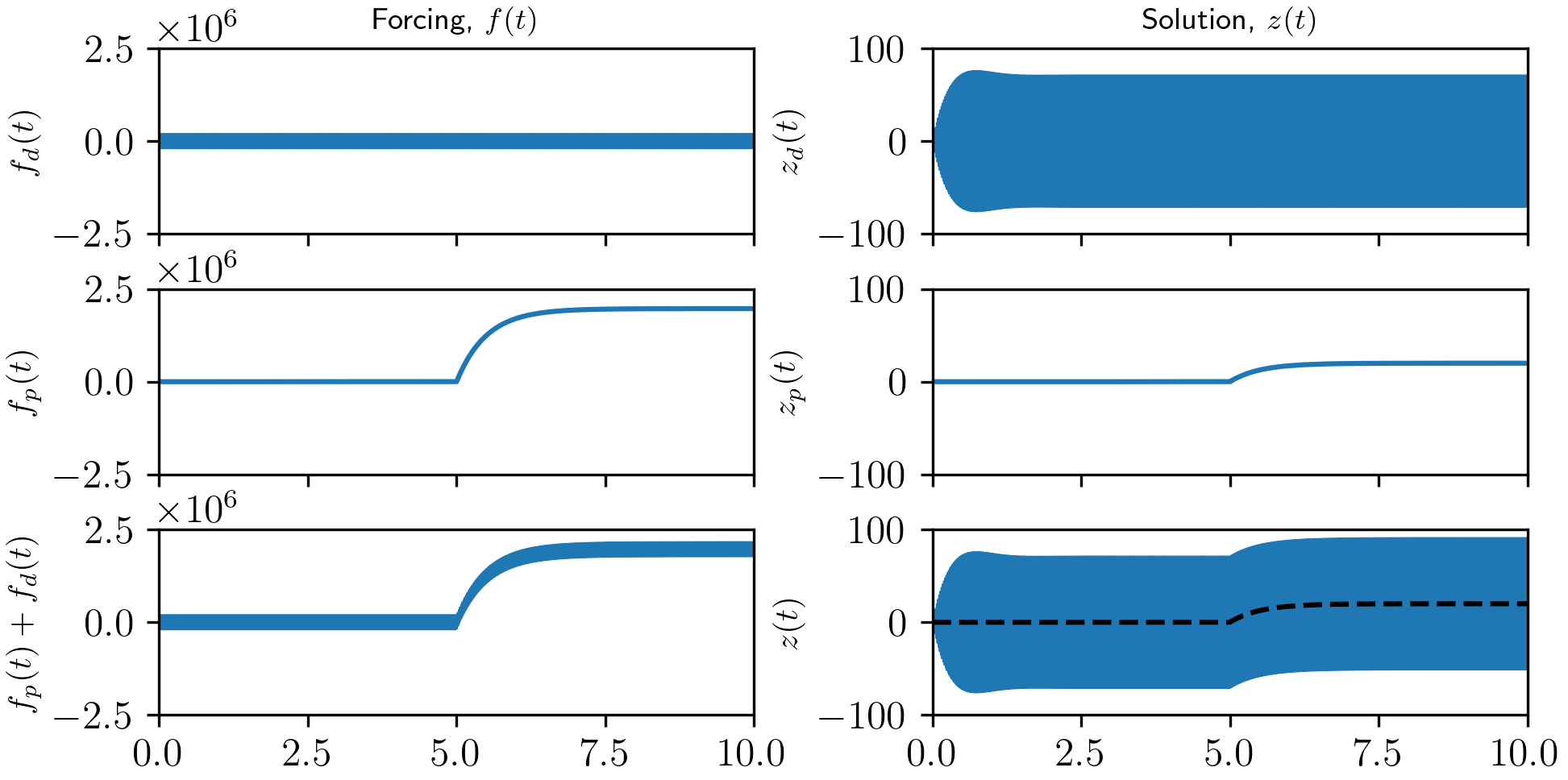}
    \caption{\textbf{Harmonic oscillator solutions with various forcings.} The left column shows the applied forcing, and the right column shows the solution to the harmonic oscillator assuming it starts from rest. The top row shows a periodic driving force, the middle row shows a non-periodic perturbation, and the bottom row shows a superposition of the two forcings.}
    \label{fig:oscillator-solutions}
\end{figure}

Having demonstrated that NFMD offers material advantages over HHT methods for TFA, both in terms of mode extraction in noisy signals and for reacting to sharp changes in instantaneous parameters, we next discuss applications to realistic systems.


\begin{figure}[t]
    \centering
    \includegraphics[width=15cm]{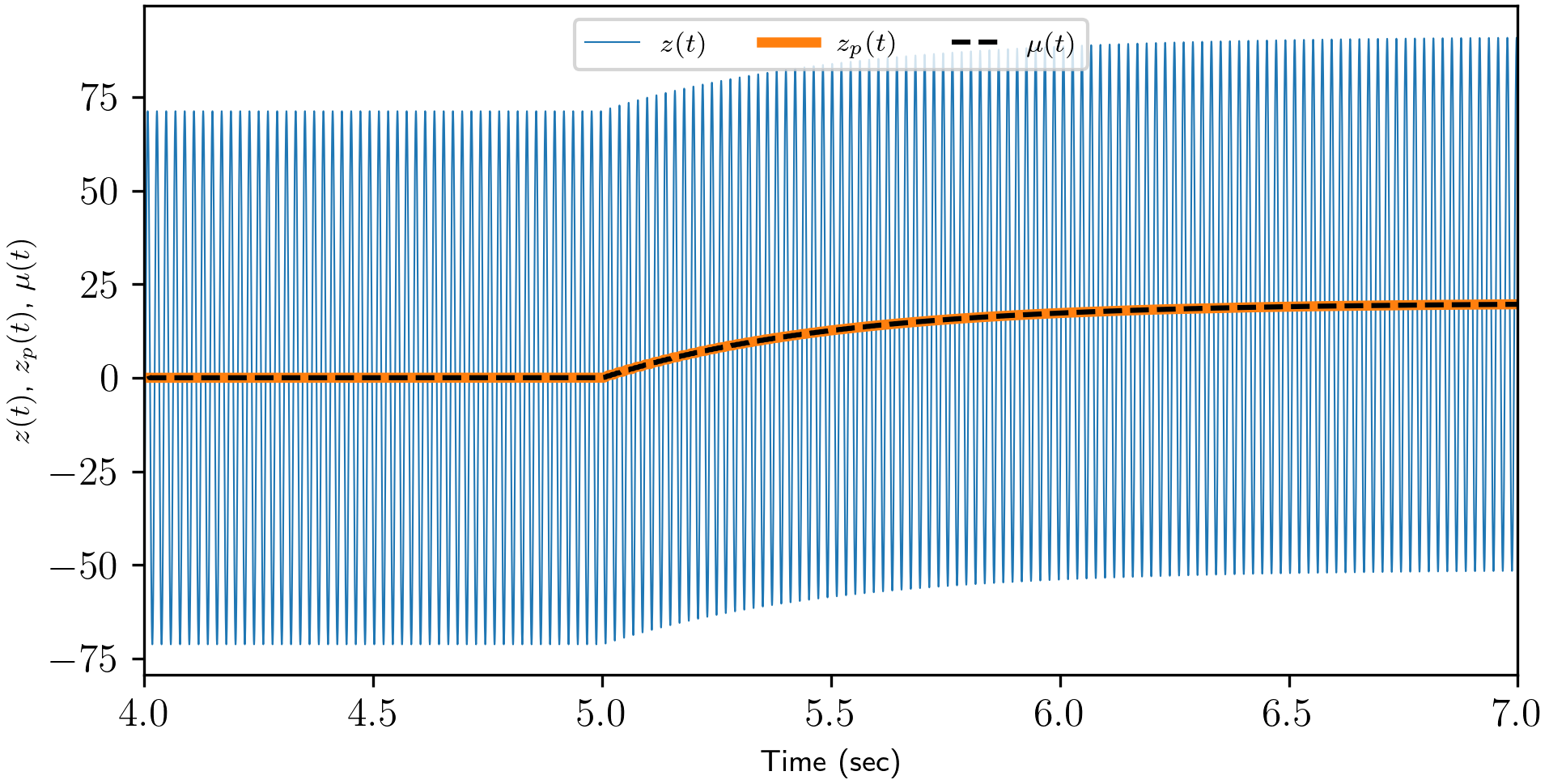}
    \vspace*{-.1in}
    \caption{\textbf{The signal mean (dashed) of the perturbed, periodically driven oscillator matches the solution of the perturbation-only oscillator (orange).} This enables NFMD to directly probe non-oscillatory driving forces applied to oscillators by accurately recovering the signal mean of a nonstationary signal.}
    \label{fig:x-xp-mean}
\end{figure}
\begin{figure}[t]
    \centering
    \includegraphics[width=15cm]{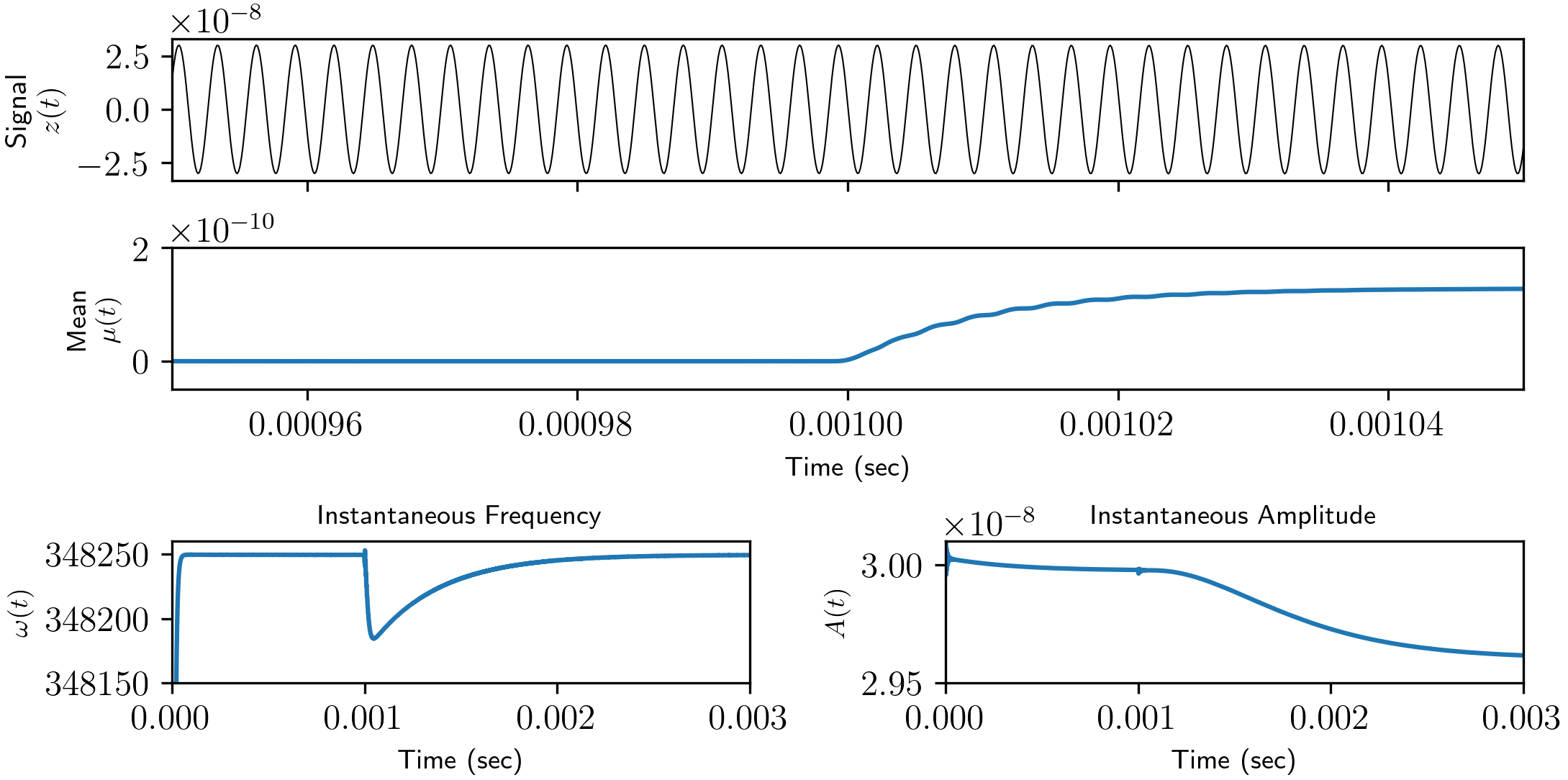}
    \vspace*{-.1in}
    \caption{\textbf{Decomposition of forced oscillator signal by NFMD.} The NFMD decomposes a harmonic oscillator forced with a periodic forcing function and non-periodic perturbation. The instantaneous mean of the signal, $\mu(t)$, is directly correlated to the non-periodic forcing.}
    \label{fig:sim-decomp}
\end{figure}

\subsection{Applications}

The example synthetic test signals above illustrate the power of NFMD to decompose signals with abruptly changing instantaneous parameters and accurately estimate the instantaneous mean of noisy signals. One application of the instantaneous mean is estimating the non-periodic forces applied to an oscillatory system. Consider the ubiquitous harmonic oscillator
\begin{equation*}
    \Ddot{x} + 2 \beta \omega_0 \Dot{x} + \omega_0^2 = F(t)/m,
\end{equation*}
where $x$ is the position of the oscillator, $\beta$ is a damping factor, $\omega_0$ is the resonant frequency of the oscillator, $F(t)$ is an applied forcing, and $m$ is the mass of the oscillator. We consider the solution to this oscillator for three types of applied forcing functions: a periodic driving force, $F_{d}(t)$, and a non-periodic perturbation forcing, $F_p(t)$, and a combination of periodic and perturbation forcing functions, $F_d(t) + F_p(t)$.

The driving force $F_d(t)$ is of the form $\alpha e^{i \omega t}$, where $\omega$ is the frequency and $\alpha$ is the amplitude of the driving force. The perturbation forcing function has the general form
\begin{equation}
    F_p(t) =  + H(t-t') \gamma (1-e^{-(t-t')/\tau}), \label{eqn-fp}
\end{equation}
where $H$ is the Heaviside function, $t'$ is a perturbation onset time, and $\tau$ is a characteristic relaxation time constant. An oscillator with this forcing function will have a solution of the form
\begin{equation}
    x_p(t) \propto  \phi e^{i \omega t} + \psi e^{-(t-t')/\tau}, \label{eqn-zp}
\end{equation}
where $\phi$ is a prefactor determined by the parameters in the governing equation. Forcing functions and solutions for an oscillator forced by $F_d(t)$, $F_p(t)$, and $F_p(t) + F_d(t)$ are presented in Figure \ref{fig:oscillator-solutions}.

The solution to the combined case ($F(t)=F_d(t)+F_p(t)$) has an instantaneous mean that is nearly identical to the solution for the perturbation-only ($F(t)=F_p(t)$) case. This enables the NFMD signal decomposition to provide insight into the form of the non-periodic forcing applied to an oscillator by estimating the instantaneous mean of the signal. The instantaneous mean and the perturbed solution are plotted on top of the solution to the driven, perturbed solution in Figure \ref{fig:x-xp-mean}.

We confirm this approach works by using a series of numerical simulations. The same periodic driving force, $F_d(t)$, is used for all simulations while a set of different perturbation forces, $F_p(t)$, is applied to each oscillator. The perturbation forces have different relaxation times, $\tau$ in equation \ref{eqn-fp}. The relaxation times vary from $10^{-7}$ to $10^{-3}$ s. The simulated oscillators are all subjected to the combined driving force and perturbation force. The signals have added white noise with SNR$=100$.

\begin{figure}[t]
    \centering
    \includegraphics[width=15cm]{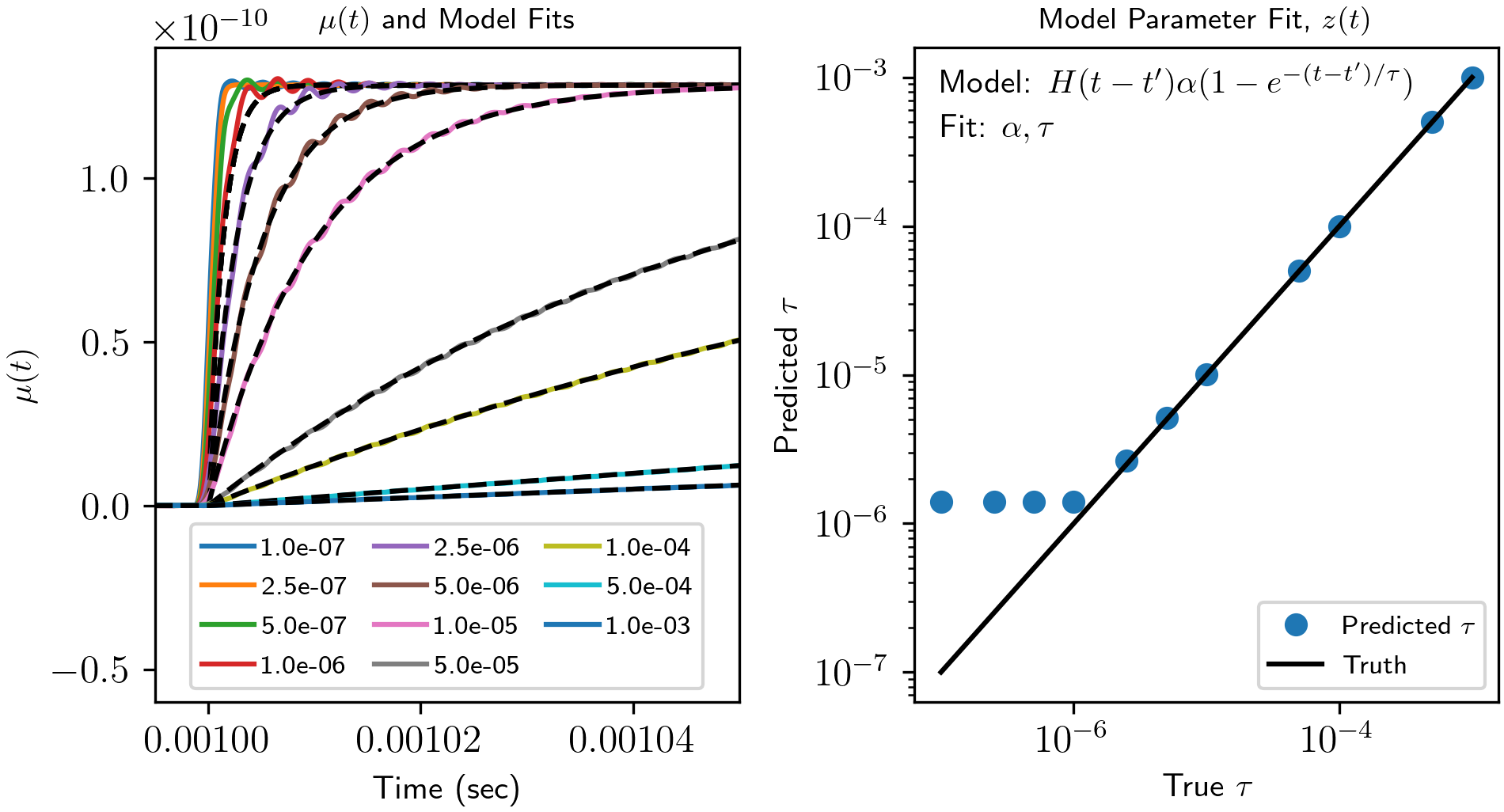}
    \caption{\textbf{Model fit comparison between instantaneous mean from NFMD and true parameters from non-periodic forcing.} The simulated oscillator is subjected to a perturbation at $t=0.001$ seconds (left panel). A model is fit to the discovered instantaneous mean mode $\mu(t)$. The time constant $\tau$ in the fit model is compared to the time constant of the perturbation force $F_p(t)$ (right panel).}
    \label{fig:sim-fits}
\end{figure}

Figure \ref{fig:sim-decomp} shows the decomposition of a single simulated oscillator. Note the time-varying mean of the signal correlates directly to the non-periodic forcing function. Figure \ref{fig:sim-fits} shows the signal means discovered with the NFMD and models fit to the instantaneous means. The model $H(t-t') \alpha (1 - \exp{}((t-t')/\tau)$ is fit to each of the instantaneous means. This model makes it possible to identify the relaxation time $\tau$ in the perturbation function. Below approximately $1 \mu s$, the estimated $\tau$ begins to flatten out. This is a result of the window size ($\xi$) used for model fitting, which is approximately a $1 \mu s$ window width.

\begin{figure}[t]
    \centering
    \includegraphics[width=15cm]{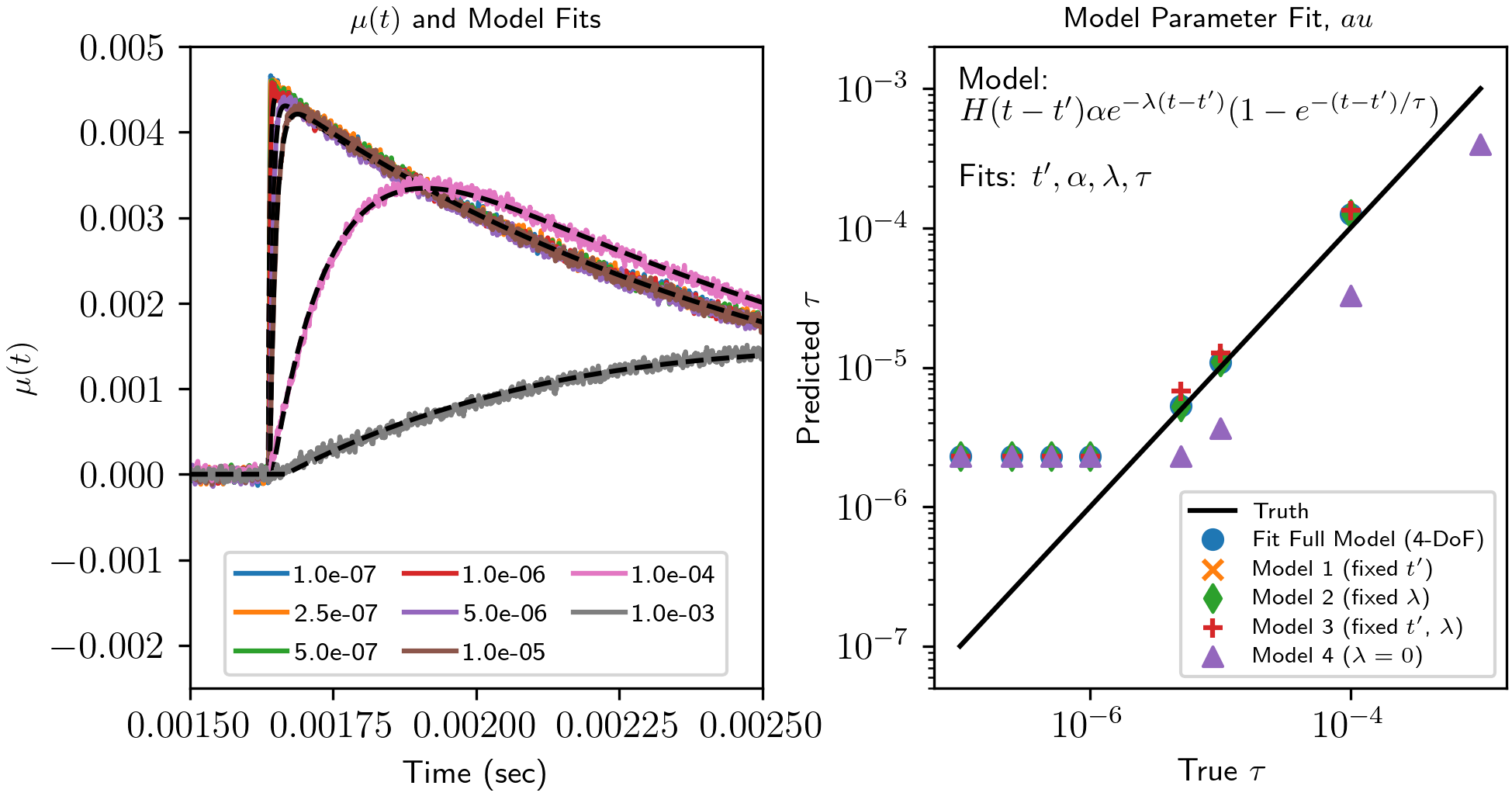}
    \caption{\textbf{NFMD decomposition of experimental trEFM control data with controlled perturbation time constants.} A model is fit which contains the time constant $\tau$ and compared with truth values.}
    \label{fig:expt-fits}
\end{figure}

In this proposed harmonic oscillator application, the NFMD can identify the instantaneous mean of a signal. The instantaneous mean of the signal is then fit to a model, which can provide insight to the form of the non-periodic perturbation applied to the system.

\subsection{Experimental Application}

We test this application on a real-world cantilever based imaging modality termed time-resolved electrostatic force microscopy (trEFM), which is an electrical modification applied to the nanoscale imaging technique of atomic force microscopy~\cite{coffey_time-resolved_2006,dwyer_lagrangian_2019,dwyer_microsecond_2017,giridharagopal_submicrosecond_2012,mascaro_review_2019}. In trEFM, a periodically driven metallic cantilever is brought into close proximity to a surface. An electric field generated by an accumulation of electrical charge on the surface imparts a force on the metallic cantilever, as well as an electrostatic force gradient~\cite{kalinin2007scanning}. The electrostatic force is typically triggered by an external stimulus, such as a voltage signal or photogenerated charge via an optical excitation source. Such methods are useful for extracting dynamic information in nanoscale measurements of photovoltaic or ionic conducting systems via the effect of the force on the cantilever's resonance frequency. In a typical trEFM experiment, the desired outcome is extracting an unknown characteristic time constant $\tau$, usually describing the time-dependent change in the electrostatic force gradient. As a test case for NFMD, we apply a series of perturbation forces to the cantilever of the form (\ref{eqn-fp}). In this experiment, the relaxation time $\tau$ is controlled in the applied forcing, providing a set of experimental data with different, known relaxation times.

The NFMD is used to decompose the experimentally-measured cantilever signal into a single periodic mode (with instantaneous frequency near the periodic driving force) and an instantaneous mean. The instantaneous mean can then be fit to a model for the forcing term. The model we used is
\begin{equation*}
    \mu(t) = H(t-t') \alpha \exp{}(-\lambda(t-t')) (1-\exp{}((t-t')/\tau),
\end{equation*}
where $H(t-t')$ is the Heaviside function centered at $t-t'$, $t$ is the time, $t'$ is the perturbation onset time, $\alpha$ is an amplitude constant, $\lambda$ is a decay constant, and $\tau$ is the relaxation time. The term with the decay constant $\lambda$ was included in this model to fit a recurring pattern in the data that is likely a constant related to the cantilever. Five different versions of this model were fit to the data: one with a fixed perturbation time $t'=0.168 \mu$s, one with a fixed decay constant $\lambda=1080$, one with both parameters fixed, one with $\lambda$ set to zero, and one with all parameters fit by the model. Figure $\ref{fig:expt-fits}$ shows the result of fitting these models to the instantaneous means of the experimental data. 

Most of the models perform similarly, though the model with the decay constant $\lambda$ set to zero tends to underestimate the correct relaxation time. Similar to the simulation results, the same trend occurs around $\tau=1 \mu s$, where the model no longer fits the truth line and the predicted $\tau$ flattens out. Unfortunately, the size of the window is limited by the frequency of the periodic components of the input signal. In both cases, the $1 \mu$s window was the minimum window size where the NFMD decomposition successfully decomposes the signal.

It is important to compare this approach with the existing methods for trEFM time constant estimation~\cite{mascaro_review_2019}. One current method is using the instantaneous frequency vector~\cite{karatay_fast_2016}, $\omega_k(t)$, of the periodic mode and estimating the time between the perturbation onset and the next local minima in the instantaneous frequency curve. Prior research showed an empirical correlation between this time interval and the relaxation time of the perturbation. To define the particular experiment, a calibration curve is used to estimate a relaxation time given the instantaneous frequency vector. This approach is effective, and enables identification of sub-microsecond relaxation time constant~\cite{giridharagopal_submicrosecond_2012}. However, the main drawback of the method in \cite{karatay_fast_2016} is that the correlation is indirect. Therefore, the external perturbation (namely, $\tau$) is not directly learned, and the calibration curve will change based on experimental variables such as the cantilever being used with a different set of physical parameters like quality factor and spring constant. For more complicated systems where the relevant timescales are more than single exponential (modern photovoltaic systems with ionic transport and dielectric relaxation in battery materials), the lack of a defined model in this calibration curve can prove limiting~\cite{tirmzi_light-dependent_2020,giridharagopal_time-resolved_2019}. A chief advantage of the proposed NFMD approach is that the forcing function can be detected directly from experimental data via the instantaneous mean.

\section{Conclusion}

Time-frequency analysis methods are critically important in science and engineering.  In this work, we develop a data-driven approach to time-frequency analysis that helps address shortcomings of classic approaches, including the extraction of nonstationary signals with discontinuities in their behavior.  By integrating elements of modern gradient descent algorithms, the Fourier transform, multi-resolution analysis, and Bayesian spectral analysis, we can learn an interpretable Fourier mode-based model for analyzing nonstationary signals with periodic components, thus circumventing the deleterious effects normally associated with nonstationary processes and allowing for accurate identification of instantaneous frequencies and their amplitudes.  Indeed, our method is equivalent to a {\em nonstationary Fourier mode decomposition} (NFMD) for nonstationary and nonlinear temporal signals. Importantly, it produces interpretable signal decompositions that can handle signals with multiple periodic components, nonlinear phase functions, and sharp discontinuities in the phase function or periodic mode amplitudes.  The method results in a superior time-frequency analysis to the HHT for nonstationary signals, and improves both temporal and spatial resolutions compared to the STFT, thus providing a viable and broadly applicable architecture for integration into a diverse number of scientific processes.

\section*{Acknowledgements}

The authors acknowledge support from the NSF MRSEC program under a SuperSeed award (1719797). DS, SLB, and JNK would especially like to acknowledge Henning Lange for helpful discussion on the FMD decomposition algorithm. Part of this work was conducted at the Molecular Analysis Facility, a National Nanotechnology Coordinated Infrastructure site at the University of Washington which is supported in part by the National Science Foundation (grant NNCI-1542101), the University of Washington, the Molecular Engineering and Sciences Institute, the Clean Energy Institute, and the National Institutes of Health.

\begin{spacing}{.01}
	\small
    \printbibliography
\end{spacing}

\end{document}